\documentclass[showpacs,preprintnumbers,amsmath,amssymb]{revtex4-1}

\newcommand{\ket}[1]{| #1 \rangle}
\newcommand{\bra}[1]{\langle #1 |}
\newcommand{\rb}[1]{\left( #1 \right)}

\newcommand{\beq}{\begin{eqnarray}}
\newcommand{\eeq}{\end{eqnarray}}

\newcommand{\op}[2]{| #1 \rangle \langle #2 |}

\newcommand{\eq}[1]{Eq.~(\ref{#1})}

\usepackage{graphicx}
\usepackage[caption=false]{subfig}
\begin{document}

\title{Collective strong coupling in multimode cavity QED}

\author{A. Wickenbrock$^{1}$, M.  Hemmerling$^{2}$, G.R.M.  Robb$^{2}$, C.  Emary$^{3}$, and F. Renzoni$^{1}$}
\affiliation{$^{1}$Department of Physics and Astronomy, University College
London, Gower Street, London WC1E 6BT, United Kingdom}
\affiliation{$^{2}$Scottish Universities Physics Alliance (SUPA), Department of Physics, 
University of Strathclyde, 107 Rottenrow, Glasgow, G4 0NG, United Kingdom}
\affiliation{$^{3}$Institut f\"ur Theoretische Physik, Technische Universit\"at Berlin, D-10623 Berlin, Germany}

\date{\today}

\begin{abstract}
We study an atom-cavity system in which the cavity has several degenerate transverse 
modes.  Mode-resolved cavity transmission spectroscopy reveals well-resolved atom-cavity 
resonances for several cavity modes, a signature of collective strong coupling for the different modes. 
Furthermore,  the experiment shows that the cavity modes are coupled via the atomic ensemble 
contained in the cavity. The experimental observations are supported by a detailed theoretical analysis. 
The work paves the way to the use of interacting degenerate modes in  cavity-based quantum
information processing, where qubits corresponding to different cavity modes interact via
an atom shared by the two modes. Our results are also relevant to the experimental realization 
of quantum spin glasses with ultracold atoms.
\end{abstract}


\maketitle

\section{Introduction}

Cavity quantum electrodynamics (cQED)  studies the interaction of atoms with a 
quantized light field enclosed in a  cavity \cite{cqed1,cqed2,cqed3}. In the early 
days of cQED, the focus was on the study of fundamental processes in atom-light interaction, and 
the exploration of the quantum-classical interface as determined by 
decoherence processes \cite{brune}. More recently, a wealth of applications 
taking advantage of the atom-cavity coupling have been identified, from 
cavity-based quantum computing \cite{qi1,qi2,qi3,qi4,qi5} to cavity cooling of 
atoms and molecules \cite{horak,vuletic00,domokos}. Most of the experimental 
and theoretical work so far has been  devoted to single mode cavities.  For 
these cavities, the regime of strong coupling was observed  with single atoms, 
with dilute atomic samples and with Bose-Einstein condensates \cite{normal,normal_bec}.  
The inclusion of multiple cavity modes in the dynamics  is predicted to 
lead to an increase of the atom-field coupling, and to  the enhancement of several cavity 
effects, as for example the atomic enhanced entanglement in optical cavities \cite{taneichi}
or an increase in the capture range of cavity cooling \cite{vuletic01}. Furthermore this 
also opens up new schemes in quantum information processing using different cavity 
modes as qubits interacting via the atoms stored in the cavity \cite{prado},  
and is an important element in the implementation of quantum spin glasses with 
ultracold atoms \cite{glasses}. Multimode cQED was realized with single atoms  interacting with different 
polarization cavity modes  \cite{ orozco}. However, in this case the number of modes
taking part ot the process was limited to two. In order to have a larger number of modes, 
a nearly-confocal cavity characterized by a large number of degenerate modes can be used.
The strong coupling regime in such a cavity was reported in Ref.~\cite{arne},
but since the mode structure could not be resolved no multi-mode features were identified.

In this work we study the coupled atom-light dynamics in a pumped  
nearly-confocal optical cavity containing cesium atoms. The near confocality
leads to a large number of degenerate transverse modes, a  
distinguishing feature of our cQED experiment. The dynamics involves multi-atoms 
multi-modes collective states.  Mode-resolved cavity transmission 
spectroscopy, as introduced in this work, allows us to study the coupling
of the atoms to the individual modes, as well as the coupling between modes
via the atoms. Our experiment reveals well-resolved atom-cavity resonances 
for several cavity modes, a signature of collective strong coupling for the different 
modes. We demonstrate the coupling, mediated by the atoms, between different 
degenerate cavity modes. This is the key element in possible implementation of 
quantum computing in cavities  using different modes as qubits \cite{prado}. 
A theoretical and numerical analysis identifies the mechanisms behind 
our observation, and demonstrates the importance of the atomic distribution 
within the modes.

This work is organized as follows. In Section II we define our set-up, and describe the experimental
results. In Sec. III we introduce the theoretical model, and interpret our experimental 
results on atom mediated modal coupling both with analytic and numerical solutions. We then 
extend the theoretical analysis beyond the parameter space of our experiment, and describe and interpret
the general situation of a multi-peaked transmission cavity profile.  Finally, in Sec. IV Conclusions
are drawn.

\section{Experiment}

\begin{figure}[t]
 \begin{center}
 \includegraphics[width=0.4\textwidth,clip=true]{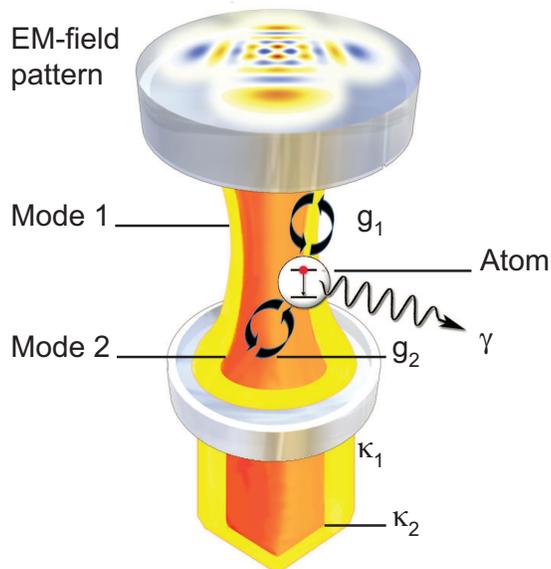}
 \end{center}
\caption{
(color online) A 2-level atom in an optical cavity interacts with different cavity modes
depending on the spatially dependent couplings. Here, two higher
order transversal modes (modes 1 and 2) and one 2-level atom exchange
excitation with rates $g_1$ and $g_2$ respectively. The coherent process
is damped by radiative coupling to the environment either via cavity
decay of each mode through the mirrors (with rates $\kappa_1$, $\kappa_2$) or
atomic polarization decay  (with rate $\gamma$).
}
\label{sketch}
\end{figure}

The details of our cavity set-up were published previously \cite{arne}. We recall here the 
essential features, and detail the additional measuring apparatus for the imaging of the 
cavity modes. The central element of the experiment is a linear, nearly confocal optical 
resonator of length $L=(11.996\pm 0.003)$ cm, as sketched in Fig. \ref{sketch}.  The relevant 
single-mode, single-atom frequencies are  
$\left(g,\kappa,\gamma\right)=2\pi\left(0.12,0.8,2.6\right)$ MHz.
The system is thus in the "bad cavity" regime $\left(g<<\kappa<\gamma\right)$. 
The beam waist $w_{00}$ of the TEM$_{00}$ cavity mode is calculated
from  the length and the mirror curvature of the cavity to be  $(127\pm 1)\mu$m. 
The cavity transmission is collimated and then imaged with an intensified CCD camera.
The resulting beam waist on the CCD chip (square pixel size: 6.45$\mu$m) is determined to
be approximately 805$\mu$m by optimising the modal decomposition with a reduced set
of modes to the empty cavity transmission coupled mostly into the 00-mode. This value is 
consistent with a beampath analysis using ray matrices and is kept fixed for all further image 
decompositions.


 Using a model incorporating spherical abberation, astigmatism and beam
front curvature the deviation from confocality $\Delta_i=L-R_i$, with $R_i$ the radius
of curvature of the mirrors along two orthogonal directions $x$ and $y$, is
derived from transmission measurements of the empty cavity. It ranges
from $\left(\Delta_x,\Delta_y\right) = \left(5.8\pm 0.7, 76\pm 4\right) \mu m$ to
$\left(\Delta_x,\Delta_y\right) = \left(22\pm 1, 94\pm 3\right) \mu m$ for
maximum and minimum piezo elongation respectively.
From the resulting frequency difference between the modes we derive the number of 
quasi-degenerate modes contained within a cavity linewidth to be 41 and 3 respectively.

One experimental cycle involves loading a magneto optical trap (MOT), placed into the
center of the cavity, from the background gas of the chamber. The loading time 
is varied between 0.1s and 2s resulting in different atom numbers between 
$8\cdot 10^4$  and $2\cdot 10^6$.  Subsequently the MOT is switched off and after
a delay of 1 ms the cavity is pumped by a linearly polarized laser beam for 1 ms,  and its transmission 
is recorded with the CCD camera. 
The relevant parameters for the pumping field are the atomic and cavity-field detunings,  
defined as $\Delta_A=\omega_P-\omega_A$  and $\Delta_C=\omega_P-\omega_C$, respectively, 
where $\omega_A$ is the atomic transition angular frequency, $\omega_C$ the cavity resonance 
frequency, which is the same for all modes as they are assumed to be degenerate, and $\omega_P$
the pump laser frequency.  Before measuring the multi-mode splitting, 
the cavity is positioned on resonance with the $F_g=4\to F_e=5$  D$_2$ line transition, i.e. 
$\omega_A=\omega_C$ and thus $\Delta_A=\Delta_C$. 
A typical experimental run starts with a probe laser detuning of  +100MHz from the 4-5 transition. 
This is then successively reduced in steps of 2MHz until a final detuning of -100MHz is reached.
For each frequency 3 transmission images are averaged.

In order to investigate the mode structure of the cavity field, the intensity profiles of 
the  measured cavity transmission are  fitted with:
\begin{align}
F(x,y) =  |\sum \alpha_{nm}  E_{nm} (x,y)|^2
\label{eq:fit}
\end{align}
where $E_{nm}(x,y)$  are  Hermite-Gaussian functions and 
$\alpha_{nm}$ the complex amplitudes. 
For a confocal cavity the mode spectrum reduces to two peaks per free spectral
range. Our experiments investigate the peak containing the subset of higher order
transversal modes degenerate with the TEM$_{q00}$ mode. Thus in our fit function 
(\ref{eq:fit}) we only consider the terms $E_{nm}$ with $n+m$ even. 
The coupling into the 
cavity TEM$_{00}$ mode was optimized by mode matching and the comparison of 
the transmission peak height of the odd  with respect to the even modes. 
For the experiment, the odd modes were suppressed by at least factor 30.

\begin{figure}[h]
\begin{center}
\includegraphics[width=0.4\textwidth,clip=true]{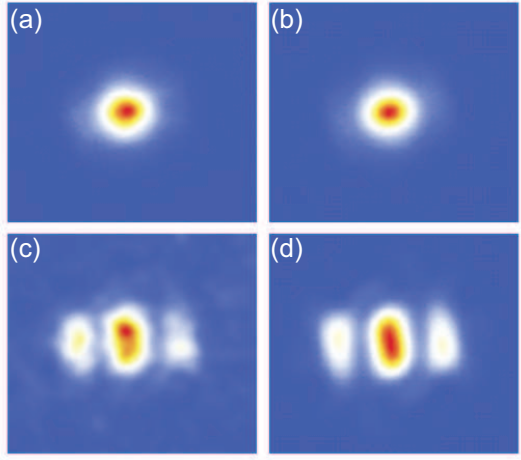}
\end{center}
\caption{
(color online) Typical images of the cavity transmission and their fits. (a) transmission through the empty cavity 
dominated by the TEM$_{00}$ mode and (b) corresponding fit;  (c) transmission through the atom-filled 
cavity with a strong TEM$_{20}$ component and (d) corresponding fit.}
\label{fig:modes} 
\end{figure}

\begin{figure}[h]
\begin{center}
\hspace{-3cm}
\includegraphics[height=3.in,clip=true]{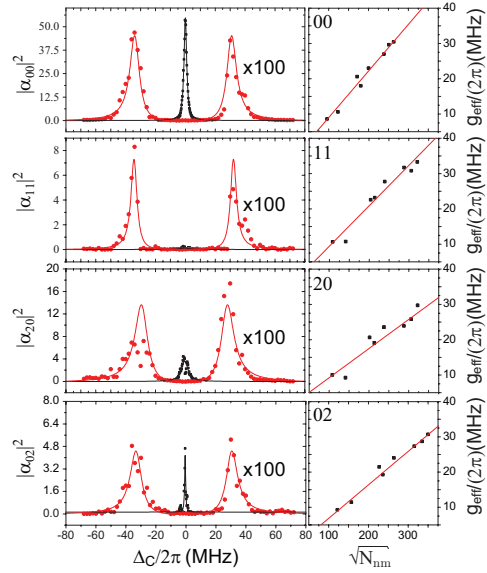}
\end{center}
\caption{
(color online) Left: Mode resolved intracavity photon number with atoms 
(red filled circles) and without
atoms (black filled squares) in the cavity for different detunings
$\Delta_C/\left(2\pi\right)$. The number of atoms in the
fundamental mode TEM$_{00}$ of the cavity as derived from fluorescence images is
$N_{00}=\left(100\pm8\right)\times10^3$. The data points
correspond to the absolute squared amplitude
$\left|\alpha_{nm}\right|^2$ of the corresponding  2D Hermite Gaussian mode
function as derived from image fits with the function Eq.~(\protect\ref{eq:fit}). The
data was then fitted with a lorentzian transmission function of an empty cavity
 and the theoretical transmission function of the coupled
cavity-atom system, as from Eq. \ref{eq:transmission}. All the displayed higher 
order transversal modes exhibit the characteristic normal mode splitting.
We stress that for one of the data set (the third figure from top, on the left) the deviation of the 
data from the fitting function  is due to experimental limitations, and does not constitute 
evidence of a further splitting.  Right: 
The effective coupling for the $nm$ mode is plotted as a function of the square root 
of the number of atoms contained in that mode. 
}
\label{fig:spec} 
\end{figure}

Typical output profiles and their fits can be seen in  Fig.~\ref{fig:modes}. A lorentzian fit to the mode resolved 
empty cavity transmission (Fig. \ref{fig:spec}, black) reveals the modal weights of the pump beam to be 
$|\alpha_{00}|^2=0.879$, $|\alpha_{20}|^2=0.054$, $|\alpha_{02}|^2=0.065$ and $|\alpha_{11}|^2=0.002$. 
The relative weight changes when the cavity is loaded with atoms. These data show clear evidence of the multi-mode dynamics of the atom-cavity system.

A detailed quantitative study is reported in Fig. \ref{fig:spec}. The mode-resolved transmission, 
both for an empty cavity and for a cavity containing cesium atoms,  is shown as a function of the 
atom-pump detuning. Normal-mode splitting is observed for several modes of the cavities.  In fact, 
although we mostly pump the TEM$_{00}$ mode, very well resolved atom-cavity resonances are 
observed also for the modes TEM$_{20}$, TEM$_{02}$, TEM$_{11}$. In particular, transmission 
for the mode TEM$_{11}$ shows no well-resolved central peak with an empty cavity, i.e 
a negligible amount of light is coupled in this mode without the atoms.  In contrast, with atoms present,
this mode shows pronounced atom-cavity resonances, which demonstrates the mediation of coupling
between different modes by the atoms. 

We fit the data for each mode with the transmission function $T(\omega_P)$ \cite{rempe}
\begin{equation}
T(\omega_P)=T_0 \left| \frac{\kappa_{eff} [\gamma+i(\omega_0-\omega_P)]}{(\omega_P-\lambda_{+})(\omega_P-\lambda_{+})}\right|^2
\label{eq:transmission}
\end{equation}
where $T_0$ is the maximum empty cavity transmission and $\lambda_{\pm} = (\omega_A\pm \Omega_0)-i(\gamma+\kappa)/2$ 
are the normal mode eigenfrequencies of the coupled system, with $\Omega_0=\sqrt{g_{\rm eff}^2-(\gamma-\kappa)^2/4}$.
The fit parameters are $T_0$,  $\kappa_{eff}$ and $g_{\rm eff}$. From our fits of the experimental data, we observed 
(see  Fig. \ref{fig:spec}, right)  that the effective coupling constant $g_{\rm eff}$ for the $nm$ mode scales as the square root of the number of
atoms contained in that mode, as expected in the case of collective strong coupling. Furthermore,
the lines were observed to broaden with the atom number, with a width scaling as $N_{nm}^{1/2}$. 
We took this feature into account by including in the fitting function an effective $\kappa_{eff}$.
As discussed below, such a broadening is a signature of the quasi-degeneracy of the effective atom-cavity couplings.

\section{Theory}

\subsection{Microscopic model}

We describe our system with an $N$-atom Tavis-Cummings model  \cite{model1,model2} extended to include $M$ cavity modes as well as an additional coherent driving term.  The Hamiltonian of the atom-cavity system reads
\begin{eqnarray}
  H = H_S + H_D = \left(  H_C + H_A +V \right) + H_D 
  \label{Hfull}
  ,
\end{eqnarray}
where $H_S$ describes the `system', i.e. the cavity, the atoms, and their interactions, and $H_D $ describes the driving.
In a frame rotating with the driving frequency, the cavity Hamiltonian reads ($\hbar=1$ throughout)
\begin{eqnarray}
  H_C &=& -\Delta_C \sum_{k=1}^M a_k^\dag a_k
  ,
\end{eqnarray}
where we have assumed that all $M$ modes are degenerate with detuning $\Delta_C$.  The atomic Hamiltonian for $N$ two-level atoms reads
\begin{eqnarray}
  H_A =  -\frac{1}{2} \Delta_A \sum_{l=1}^N \sigma_l^z
  , 
\end{eqnarray}
where $\Delta_A$ is the detuning of the atoms, assumed identical for all atoms.  The interaction between the atoms and the cavity modes in the rotating wave approximation (RWA) reads
\begin{eqnarray}
 V &=& 
  \frac{1}{2}
  \sum_{k=1}^M 
  \sum_{l=1}^N 
  \left\{
    g_{kl} a_k^\dag \sigma^-_l
    +
    g^*_{kl} a_k  \sigma^+_l
  \right\}
  \label{V}
  .
\end{eqnarray}
 
Here, $a_k^{\dagger}$ and $a_k$ are the creation and the annihilation operators 
of $k$th mode, where the index $k$ describes the complete set of modes previously 
labelled $nm$, $\sigma_l^\pm$ are the atomic raising and the lowering operators of 
$l$th atom, and $g_{kl}$ is the coupling constant for the interaction between the $k$th
mode and the $l$th atom, which depends on the 
atom via its position within the cavity.

 We drive the system with a monochromatic laser, which we assume only couples to the cavity modes.  In the RWA, we have
\begin{eqnarray}
   H_D &=& 
  \sum_{k=1}^M \eta_k 
  (a_k^\dag + a_k )
  ,
\end{eqnarray}
with $\eta_k$ the coupling strength of the laser to mode $k$.  Dissipation due to cavity decay and spontaneous emission 
with homogeneous 
rates $\kappa$ and $\gamma$ respectively, is introduced in the standard quantum master 
equation approach.

\subsection{Weak excitation model }
We will consider that the effect of the driving term $H_D$ is weak enough that it introduces at most one excitation (atomic flip or cavity photon) into the system at any given time.  This assumption allows us to restrict ourselves to the following set of relevant states:
\begin{eqnarray}
  | 0 \rangle      &:& \quad \text{no excitation (cavity vacuum, all atoms in ground state)} \nonumber\\
  | C_k \rangle  &:& \quad \text{single-photon excitation in cavity mode}~k\nonumber\\
  | A_l \rangle   &:& \quad \text{excitation of atom}~l\nonumber\\
\end{eqnarray}
In this basis, the component Hamiltonians of Eq.~\ref{Hfull} (minus an offset) read
\begin{eqnarray}
  \label{weakmodel}
  H_C &=& - \Delta_C \sum_{k=1}^M   | C_k \rangle \langle C_k |~;\\
  H_A &=&  -\Delta_A \sum_{l=1}^N  | A_l \rangle \langle A_l |~;  \\
  V &=& 
  \frac{1}{2}\sum_{k=1}^M \sum_{l=1}^N
  \left(
    g_{kl}  {| C_k \rangle \langle A_l |} +g^*_{kl}  {| A_l \rangle \langle C_k |}
  \right)
  ;
  \\
  H_D &=& 
  \sum_{k=1}^M \eta_k \left(  | C_k \rangle \langle 0 | +   | 0 \rangle \langle C_k |      \right)  
  .
\end{eqnarray}

\subsection{Singular value decomposition of the coupling matrix}

Let us write the coupling constants as the $M\times N$ matrix $G$ with elements $(G)_{kl}=g_{kl}$.  The singular value decomposition (SVD) of $G$ reads
\begin{eqnarray}
  G = U \Lambda W^\dag
  ,
\end{eqnarray}
where $U$ is an $M\times M$ unitary matrix, $W$ an $N\times N$ unitary matrix, and $\Lambda$ is an $M\times N$ rectangular diagonal matrix.  The  non-zero diagonal elements of $\Lambda$ are the singular values of $G$, which we denote $\lambda_j$.  The singular values are real, positive and equal to the non-zero eigenvalues of the matrix  $\Gamma \equiv \sqrt{G G^\dag}$ and the number of singular values is equal to the rank,  $R_\Gamma$, of this matrix.  Since, in our system, the number of modes is smaller than the number of atoms, $M < N$, the coupling matrix $G$ can have at most $M$ singular values, such that $R_\Gamma \le M$.

Using this SVD we can rewrite the interaction as
\begin{eqnarray}
  V &=& 
  \frac{1}{2} \sum_{k=1}^M \sum_{l=1}^N \sum_{j=1}^{R_\Gamma}
  U_{kj}\; \lambda_j \; (W^\dag)_{jl} \op{C_k}{A_l}  + \text{H. c.}
  \nonumber
  \\
  &=& 
  \frac{1}{2}  \sum_{j=1}^{R_\Gamma}
  \rb{ \sum_{k=1}^M  U_{kj} \ket{C_k}}
  \lambda_j
  \rb{\sum_{l=1}^N  (W^\dag)_{jl}\bra{A_l}}
  + \text{H. c.}
  .
\end{eqnarray}
We then define the {\em collective} cavity and atomic states:
\begin{eqnarray}
   \ket{\widetilde{C}_j} &=& \sum_{k=1}^M  U_{kj} \ket{C_k} \quad \quad \mathrm{(cavity)}
   \label{colstateC}\\
    \ket{\widetilde{A}_j} &=& \sum_{l=1}^N W_{lj}\ket{A_l}\quad \quad \mathrm{(atomic)}
    \label{colstateA}
\end{eqnarray}
and obtain
\begin{eqnarray}
  V &=&  \frac{1}{2}  \sum_{j=1}^{R_\Gamma} \lambda_j 
  \left(
   | \widetilde{C}_j \rangle \langle \widetilde{A}_j | 
    +
   | \widetilde{A}_j \rangle \langle \widetilde{C}_j | 
  \right).
\end{eqnarray}
In this new basis, the remaining parts of the system Hamiltonian read
\begin{eqnarray}
  H_C &=& - \Delta_C \sum_{k=1}^M   | \widetilde{C}_k \rangle \langle \widetilde{C}_k |~;\\
  H_A &=&  - \Delta_A \sum_{k=1}^M  | \widetilde{A}_k \rangle \langle \widetilde{A}_k | ~.
\end{eqnarray}
The atomic part does, in fact, contain other terms ($N-M$ of them), but these are all decoupled from the cavity and play no further role. 
Taken together, then, we can write the system Hamiltonian as $H_S =  h_0 +  \sum_{j=1}^{R_\Gamma} h_j$ where
\begin{eqnarray}
  h_j &= &
    -\Delta_C \op{\widetilde{C}_j}{\widetilde{C}_j}
    - \Delta_A  \op{\widetilde{A}_j}{\widetilde{A}_j}
    + \frac{1}{2}  \lambda_j 
    \rb{
      \op{\widetilde{C}_j}{\widetilde{A}_j} 
      +
      \op{\widetilde{A}_j}{\widetilde{C}_j}
    }
  ,
\end{eqnarray}
describes the coupling between the $j$th collective atom and cavity modes, 
and 
\beq
  h_0 = \sum_{j= R_\Gamma+1}^M
  -\Delta_C \op{\widetilde{C}_j}{\widetilde{C}_j}
  - 
  \Delta_A \op{\widetilde{A}_j}{\widetilde{A}_j}
  ,
\eeq
describes the remaining uncoupled elements that arise from a rank-deficient $\Gamma$ matrix. In an obvious matrix representation, the coupled Hamiltonian can be written
\beq
  h_j &= &
  \left(
  \begin{array}{cc}
    -\Delta_C & \frac{1}{2}\lambda_j \\
    \frac{1}{2}\lambda_j  & -\Delta_A
  \end{array}
  \right)
  .
\end{eqnarray}
Thus, in the weak-excitation limit, we see that we obtain a set of $R_\Gamma$ independent two-level coupled atom-cavity systems.  These two-level system have identical detunings, but each has its own effective coupling strength, $\lambda_j$.  On resonance, $\Delta_A=\Delta_C =0$, the splitting of the two states in the $j$th system is simply $\lambda_j$.

In the collective basis the driving term reads
\beq
  H_D &=& 
  \sum_{j=1}^M  \rb{\tilde{\eta}_j \op{\widetilde{C}_j}{0} +\tilde{\eta}^*_j \op{0}{\widetilde{C}_j}}  
  ,
\eeq
with the new driving amplitudes 
$
  \tilde{\eta}_j =\sum_{k=1}^M U_{jk}^*\eta_k
$.
Thus we see that, since the SVD mixes all cavity modes together (for a generic coupling matrix $G$), pumping just one of the original modes pumps all the collective modes.

\subsection{Effective coupling strengths and the $\Gamma$-matrix}

The cavity transmission spectrum is determined by the eigenvalues $\lambda_j$ of the matrix $\Gamma = \sqrt{GG^{\dag}}$.
The elements of $G$ are the microscopic atom-cavity coupling constants $g_{kl}$, which depend on the atomic position: $g_{kl} = g_k(\mathbf{r}_l)$.  The eigenvalues $\lambda_j$, and thus the cavity transmission spectrum, are therefore expected to depend upon the distribution of the atoms within cavity.
In the continuum limit, the atomic positions may be described by the continuous density distribution $\rho_A({\bf r})$. The matrix product in $GG^{\dagger}$ is then transformed into an average over this atomic distribution
\begin{equation}
  \rb{GG^{\dag}}_{kk'} 
  \to 
  \int d^3{\bf r} 
     \rho_A(\mathbf{r})  g_k(\mathbf{r}) g^*_{k'}(\mathbf{r}) 
  \equiv
  \rb{\langle G G^{\dag} \rangle_A}_{kk'}
  .
\end{equation}

We describe the cavity modes with the set of Hermite-Gaussian polynomials $u_{nm}$, which are orthogonal and normalized so as to have a fixed volume 
integral equal to the volume of the 00-mode $V_{00} = L\pi w_{00}^2/4$, with $L$ the cavity length and $w_{00}$ the
waist of the 00-mode. Replacing the single mode index $k$ with the double index $(nm)$ of the Hermite-Gaussian polynomials, the coupling constant
of the $nm$-mode can be written as
\begin{equation}
  g_{(nm)}({\bf r}) = \sqrt{
  \frac{\mu^2 \omega_C}{2\hbar \epsilon_0 V_{00}} 
  }
  u_{nm}({\bf r})
  ,
\end{equation}
where $\mu$ is the atomic dipole moment, $\omega_C$ the cavity resonance frequency and $\epsilon_0$ the vacuum permittivity.
Then the matrix $GG^{\dag}$ can then be written as
\begin{equation}
\langle GG^{\dag} \rangle_A = \frac{\mu^2 \omega_C}{2\hbar \epsilon_0 V_{00}}  \left( 
\begin{array}{cccc}
\langle u_{00}^2         \rangle_A & \langle u_{00} u_{11}  \rangle_A & \langle u_{00}u_{20}  \rangle_A & ..... \\
\langle u_{11}u_{00}  \rangle_A & \langle u_{11}^2          \rangle_A & \langle ...  \rangle_A & ..... \\
            ...                                &              ...                                  & \langle u_{20}^2  \rangle_A & ..... \\
            ...                                &              ...                                   &   ...  & ..... 
\end{array}
.
\right)
\end{equation}
In the limiting cases of uniform and delta-peaked atomic distributions, it is possible to perform the average over atomic degrees of freedom explicitly and find analytic expressions for the effective coupling strengths $\lambda_j$, as we now discuss.

\subsubsection{Uniform atomic distribution}

We consider first the case of a uniform atomic distribution
\begin{equation}
  \rho_A = \frac{N}{V}~.
\end{equation}
The matrix $\langle GG^{\dag} \rangle_A$ becomes diagonal due to the orthogonality of the Hermite-Gaussian polynomials:
\begin{equation}
  \rb{\langle G G^{\dag} \rangle_A }_{kk'}
  = 
  \frac{\mu^2 \omega_C}{2\hbar \epsilon_0 V_{00}} \rho_A V_{00} \delta_{kk'}
  =
  \frac{\mu^2 \omega_C}{2\hbar \epsilon_0} \rho_A \delta_{kk'}
  .
\end{equation}
The eigenvalues of the matrix $\Gamma$ are then degenerate, with the value
\begin{equation}
  \lambda_k =  \sqrt{ \frac{\mu^2 \omega_C}{2\hbar \epsilon_0} \rho_A }
  ~,
\end{equation}
and the matrix $\Gamma$ is of full rank, $R_\Gamma = M$.
The effective couplings $\lambda_k$ do not show any dependence on the number of modes $M$, and thus, with a uniform atomic distribution, the multiple degenerate cavity modes do not lead to an enhancement of the coupling.  Rather, there exist $M$ such split systems with the same coupling. It should be noted that the cavity states involved in each of these splittings are mixtures of all of the original modes, see \eq{colstateC}.

\subsubsection{Delta-peaked atomic distributions}

We consider now the opposite case of an atomic distribution with all atoms localised in the centre of the cavity, $\mathbf{r} = \mathbf{0}$, and we take the density function to be a delta function:
\begin{equation}
  \rho_A ({\bf r}) = N \delta({\bf r})~.
\end{equation}
The matrix $GG^{\dag}$ is readily calculated as 
\begin{equation}
  \langle GG^{\dag} \rangle_A = N  \left( 
  \begin{array}{cccc}
   g_{(00)}^2               &   0         &  g_{(00)} g_{(22)}   & .... \\
   0                             &   0         &             0                                           & ..... \\
   g_{(00)} g_{(20)}       &   ...       &  g_{(20)}^2   & ..... \\
             ...                 &   ...       &            ...                           & ..... 
  \end{array}
  \right)
  ,
\end{equation}
where all the coupling constants $g_{(nm)}$ are evaluated at ${\bf r}={\bf 0}$. The matrix $GG^{\dag}$ has a single non-zero eigenvalue 
($\lambda_1^2$, say) given by
\begin{equation}
  \lambda_1^2 = N \sum_{k=1}^{M} g_k^2
  ,
\end{equation}
and all the others equal to zero  ($R_\Gamma =1$). In this case the presence of $M$ degenerate modes leads to an enhancement of the atom-cavity coupling,
in contrast to the case analyzed previously. For the case $g_k \sim g_{00}$, we recover the known result \cite{papa} $\lambda_1 \sim g_{00}\sqrt{NM}$ such that the single 
effective coupling scales with both the square root of the number of both atoms and cavity modes.

That $\Gamma$ is rank deficient in this case implies that the cavity transmission will display a peak for zero detuning,  provided that any of the decoupled collective
modes (i.e. those corresponding to a zero eigenvalue) are pumped. This, as well as the case of more generic atomic distributions, will be further analyzed with
the help of numerical simulations.


\subsection{Quantum master equation}
To include the effects of cavity losses and spontaneous emission we study the
Liouville-von-Neumann equation for the atom-cavity density matrix $\rho$, which reads
\begin{equation}
  \frac{d}{dt}\rho = -i \left[H, \rho\right] 
  +\mathcal{L}_\mathrm{loss}[\rho]
  +  \mathcal{L}_\mathrm{spon}[\rho].
\end{equation}
Here
\begin{equation}
  \mathcal{L}_\mathrm{loss}[\rho] = 
  \frac{1}{2}\sum_{m=1}^M 
  \kappa_m 
  \left\{
    2 a_m \rho a_m^\dag 
    - a_m^\dag a_m \rho 
    - \rho a_m^\dag a_m  
  \right\}
\end{equation}
describes the loss at rate $\kappa_m$ from cavity-mode $m$, and
\beq
  \mathcal{L}_\mathrm{spon}[\rho] = 
  \frac{1}{2}\sum_{n=1}^N 
  \gamma_n
  \left\{
    2 \sigma_n^- \rho \sigma_n^+ 
    - \sigma_n^+ \sigma_n^- \rho 
    - \rho \sigma_n^+ \sigma_n^-
  \right\}
\eeq
describes the spontanenous emission from the atoms at rate $\gamma$. For simplicity, in calculations 
we assume that all modes decay at the same rate $\kappa_m = \kappa;\forall m$. 

Our simulations follow the dynamical evolution of N 2-level atoms and M cavity modes where 
the atoms are randomly spatially distributed and the modes are Hermite-Gaussian $TEM_{nm}$ modes.
By simulating single trajectories of the $N$ atom and $M$ mode system, we
are able to study the dynamics of a cavity with four modes, and containing 
up to $1.6\times 10^4$ atoms. The experimentally observed four strongest modes were
considered in our simulations.
The atom-cavity couplings $g_{ka}$ are determined by assuming an atomic distribution
within the cavity corresponding to the experimental one. We generate  
$N$ random positions according to the atomic distribution, and then calculate the 
$N \times M$ couplings for the $N$ atoms coupled to the $M$ modes. 
Pumping only one mode (TEM$_{00}$), the evolution of the photon number in each mode,
$|\alpha_{nm}|^2$, shows the coupling between modes induced via the atom-cavity interaction. 

Our simulation deals with a number of atoms smaller than the one considered experimentally,
and a larger single atom coupling constant for computational simplicity.
We verified numerically that the same results produced for a small number of atoms and large
single-atom coupling apply for large number of atoms and weak single-atom coupling, provided
that the system is in the collective strong coupling regime.

\subsubsection{Mode-mode coupling}

\begin{figure}[t]
\begin{center}
\hspace{-3cm}
\includegraphics[width=0.5\textwidth,clip=true]{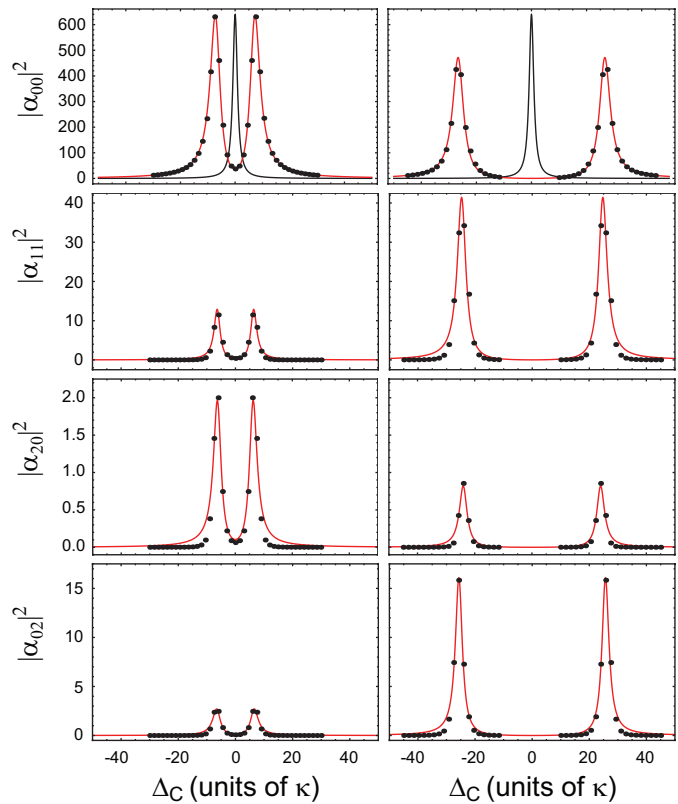}
\end{center}
\caption{
(color online) Numerical results for the intracavity photon number as a function 
of  $\Delta_C$ for the the case of 1000 (left) and 16000 (right) atoms gaussian 
distributed in a realistic 3D multi-mode cavity field with the available TEM modes 00,20,02 and 11.
The (black) dots are the actual numerical results from the simulation, while the
solid (red) lines are the best fits with Eq. \ref{eq:transmission}.
Dashed (black) lines are results for an empty cavity, reported for comparison.
The parameters of the calculation are as follows. The maximum value for the coupling 
constant for the 00-mode is $g=\kappa$, with the individual couplings for the different atoms
calculated depending on their position (see text). $\gamma = 3.25 \kappa$.
The mode pumped is the TEM$_{00}$ mode with $\eta_{00} = 0.1\kappa$ and the
displayed steady state photon number is multiplied by $10^6$. The atomic
distribution is similar to the experiment following a cloud with
gaussian width $\sigma_z=1.25 w_0$  and $\sigma_r=1.6 \sigma_z$ ,
where $w_0$ is the waist of the cavity field and $z$ is perpendicular to the cavity axis. 
} 
\label{Figure1}
\end{figure}

We first consider an atomic distribution closely representing the experimental one, and investigate the coupling between the modes mediated by the 
atomic sample. Our numerical results for the  steady-state intracavity photon number for the different modes are reported in Fig. 
\ref{Figure1}. These results  show that there is an effective atom-mediated coupling between the modes: by pumping only one mode of the cavity (mode 1) 
the presence of the atoms leads to scattering also into the other modes, and a cavity field is built up also for the modes not pumped by the laser. This is 
in agreement with our experimental results.

We notice that in general we would expect to see a multi-peaked mode-profile for each cavity optical mode
with splittings corresponding to the different eigenvalues of the $\Gamma$-matrix.  However, both experiment
and numerical simulations show just two peaks in the transmission spectrum of each mode, and no 
features at resonance with the bare cavity frequency. 

The observed two peaked structure can be understood by studying the eigenvalues of the $\Gamma$-
 matrix as derived from the atomic density function. In the limit of a spatially uniform atom distribution the 
$\Gamma$-matrix becomes diagonal and its eigenvalues degenerate, consistent with the observed 
two-peak structure.   

In our system the atomic distribution is not exactly uniform, but large enough to make the effective  
couplings $\lambda_k$  quasi-degenerate, hence the absence of a multi-peaked 
structure in the cavity transmission spectrum for the different modes.

\subsubsection{Line broadening}

\begin{figure}[tb]
 \begin{center}
 \includegraphics[width=0.4\textwidth]{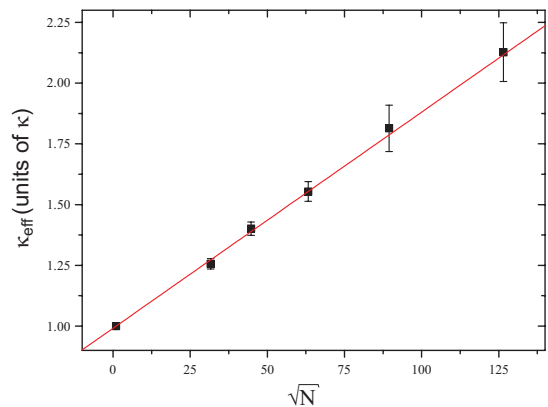}
 \end{center}
\caption{
(color online) Value for the parameter $\kappa_{\rm eff}$ as derived from fitting the numerically simulated transmission data of the TEM$_{00}$ mode data with eq. 2 as a function of the number of atoms $N$. The atoms are randomly distributed in a 3D Gaussian centred with the centre of the cavity, and with width $\sigma_x=1.3w_0$ and $\sigma_y=\sigma_z=1.6\sigma_x$. The  cavity has four transverse modes: 00, 11,20,02. The mode pumped is the 00-mode with $\eta_{00} = 0.1\kappa$. Since the eigenvalues of the system are quasi degenerate the resulting effective $\kappa$ scales linearly with $\sqrt{N}$, where $N$ is the total atom number in the distribution.}
\label{figure_2_supp}
\end{figure}

While the quasi-degeneracy of the effective couplings does not allow the resolution of a multi-peak structure, it leads to a broadening of the cavity transmission peaks with increasing 
number of atoms. As the eigenvalues scale with $N^{1/2}$, the cavity 
transmission peaks should broaden with the same scaling. We run
numerical simulations to analyze the width of the cavity transmission 
spectrum as a function of the number of atoms in the cavity. In our calculations we
selected the parameters corresponding to our experiment, with results in 
Fig.~\ref{figure_2_supp}. The simulations show the broadening of the width of the
 resonance with the dependence $N^{1/2}$ as observed in our experiment. This can be traced back to 
quasi-degenerate eigenvalues $\lambda_j$ which scale as $N^{1/2}$, thus leading to a broadening of the transmission peaks with the
same dependence on the atom number. That is, the transmission peak is actually made up of the several near degenerate modes,    i.e. the experimentally observed line broadening is actually an overlap of several lines    which are each too broad to be resolved.  This behaviour holds as long as the spreading $\Delta \lambda$ of the eigenvalues is much smaller 
than the empty cavity linewidth $\kappa (0)$. Otherwise the spectrum can separate in several peaks.

\subsubsection{Multi-peaked cavity transmission}

We generalize our analysis by considering atomic distributions different from the one corresponding to our experiment. This allows us to explore multi-peaked 
cavity transmission profiles.

\begin{figure}[tb]
 \begin{center}
 \includegraphics[width=.5\textwidth]{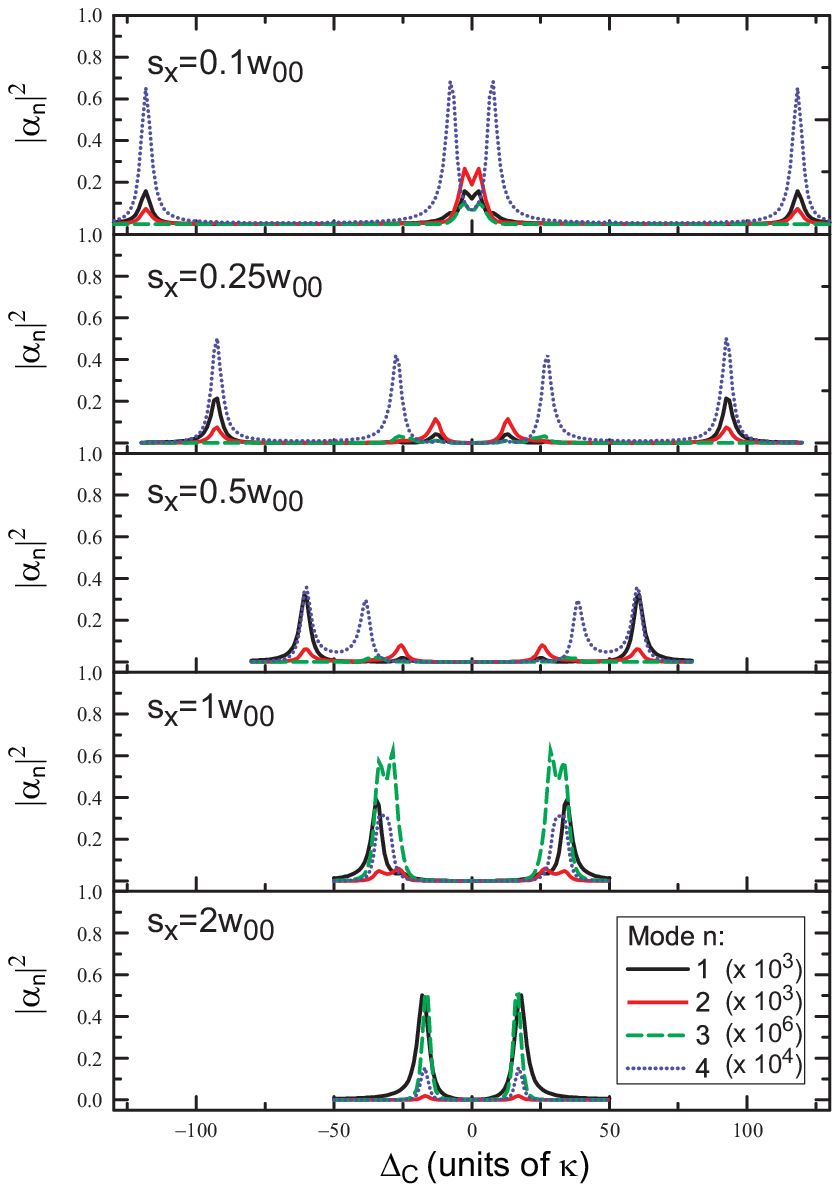}
 \end{center}
\caption{
(color online) Numerical simulations for the cavity transmission spectroscopy for different spatial extensions of the atomic cloud.
 All the data sets are for a sample of 16000 atoms, and a cavity with 4 degenerate modes.  The atoms are 3D Gaussian distributed with width $(s_x,2s_x,2s_x)$, expressed in the figure
in terms of the fundamental Gaussian cavity mode waist $w_{00}$. The different data plots refer to different widths $s_x$ of the atomic distribution.
The mode pumped is the 00-mode with $\eta_{00} = 0.1\kappa$.}
\label{fig_1_supp}
\end{figure}

Our simulations examine the structure of the cavity transmission spectrum for different atomic distributions, with results as those shown in Fig.~\ref{fig_1_supp}. 
In all our simulations, we pump only the fundamental mode (mode 00).

For a broad atomic distribution (bottom panel) only two peaks, symmetrically displaced with respect to the origin, are present in the mode spectrum. This is consistent
with our analytic results for an uniform distribution. 
For narrower atomic distributions the mode structure is more complicated, in agreement with the fact that the eigenvalues $\lambda_j$ are not degenerate 
any more. 
Finally, for atomic distributions more and more localized in the middle of the cavity, the mode structure simplifies again: a central peak appears, together 
with two symmetrically displaced resonances. This is in agreement with  our analytic results for a delta-peaked atomic distribution.

\section{Conclusions}

In conclusion, in this work we studied an atom-cavity system, in which the cavity 
has several degenerate transverse modes. This was obtained by using a nearly 
confocal cavity.  Experiments were performed in the collective strong coupling regime. 
Mode-resolved cavity transmission spectroscopy revealed well-resolved atom-cavity 
resonances for several cavity modes. In agreement with our theoretical model, the 
experiment shows that the cavity modes are coupled via the atomic ensemble contained in the cavity.

The present work is of relevance to a number of applications, from cavity cooling of atoms
and molecules, where the use of nearly-confocal cavities is predicted to enhance the cooling, to 
cavity quantum information processing, where atom-mediated interaction between degenerate modes 
may allow novel implementation of quantum gates as well the production of highly 
entangled states of different cavity mode states. Finally, our results are also relevant to 
the experimental realization of quantum spin glasses with ultracold atoms.

\acknowledgements

This work is supported by EPSRC (grant EP/H049231/1), and the Royal Society.


\begin{thebibliography}{99}
\bibitem{cqed1}
{\it Cavity Quantum Electrodynamics}, P. Berman, Ed. (Academic Press, Boston, MA, 1994).
\bibitem{cqed2}
H. Mabuchi,  and A.C. Doherty, Science {\bf 298}, 1372 (2002).
\bibitem{cqed3}
J.M. Raimond,  M. Brune, and S. Haroche,  Reviews of Modern Physics {\bf 73}, 565 (2001).
\bibitem{brune}
M. Brune, E. Hagley, J. Dreyer, X. Ma\^itre, A. Maali, C. Wunderlich, J. M. Raimond, and S. Haroche,
Phys.Rev.Lett. {\bf 77}, 4887 (1996).
\bibitem{qi1}
T. Pellizzari, S.A. Gardiner, J.I. Cirac, and P. Zoller,  Phys.  Rev. Lett. {\bf 75},
3788 (1995).
\bibitem{qi2}  S.J. van Enk, J.I. Cirac, and P. Zoller,  Phys.  Rev. Lett. {\bf 79}, 
5178 (1997).
\bibitem{qi3}
J.  Pachos, and H. Walther,  Phys. Rev. Lett. {\bf 89}, 187903 (2002).
\bibitem{qi4}
L.M.  Duan, A. Kuzmich, and H.J. Kimble,  Phys.  Rev.  A {\bf 67}, 032305 (2003);
\bibitem{qi5}
Q.A. Turchette, C.J. Hood, W. Lange, H. Mabuchi, and H.J. Kimble,  
Phys. Rev. Lett. {\bf 75}, 4710 (1995).
\bibitem{horak}
P. Horak, G. Hechenblaikner, K. M. Gheri, H. Stecher, and H. Ritsch,
Phys. Rev. Lett. {\bf 79}, 4974  (1997) .
\bibitem{vuletic00}
V. Vuletic and S. Chu, Phys. Rev. Lett. {\bf 84}, 3787 (2000)
\bibitem{domokos}
P. Domokos and H. Ritsch, J. Opt. Soc. Am. {\bf B 20}, 1098 (2003).
\bibitem{normal}
M.G. Raizen, R.J. Thompson, R.J. Brecha, H.J. Kimble, and H.J. Carmichael,
Phys. Rev. Lett. {\bf 63}, 240 (1989);
R.J. Thompson, G. Rempe, and H.J. Kimble, Phys. Rev. Lett. {\bf 68}, 1132 (1992).;
A.K. Tuchman, R.  Long,  G. Vrijsen, J. Boudet, J. Lee,  and M.A. Kasevich, 
Phys. Rev. A {\bf 74}, 053821 (2006).
\bibitem{normal_bec}
Y. Colombe, T. Steinmetz, G. Dubois, F. Linke, D. Hunger, and J. Reichel,
Nature {\bf 450}, 272 (2007);
F. Brennecke, T. Donner, S. Ritter, T. Bourdel, M. K\"ohl, and T. Esslinger,
Nature {\bf 450}, 268 (2007).
\bibitem{taneichi}
T. Taneichi and  T. Kobayashi, Chem. Phys. Lett. {\bf 378}, 576 (2003).
\bibitem{vuletic01}
V. Vuleti\'c, H.W. Chan, and A.T. Black, Phys. Rev. A {\bf 64}, 033405 (2001). 
\bibitem{prado}
F.O. Prado, F. S. Luiz, J. M. Villas-B\^oas, A. M. Alcalde, E. I. Duzzioni, and L. Sanz,
Phys. Rev. A {\bf 84}, 053839 (2011);
Y. Dong, X. Zou, S. Zhang, S. Yang, C. Li, and G. Guo, J. Mod. Opt. {\bf 56}, 1230 (2009). 
\bibitem{glasses}
P. Strack and S. Sachdev, Phys. Rev. Lett. {\bf 107}, 277202 (2011).
\bibitem{orozco}
M.L. Terraciano, R. Olson Knell, D.G. Norris, J. Jing, A. Fern\'andez, and L.A. Orozco,
Nat. Phys. {\bf 480}  (2009).
\bibitem{arne}
A. Wickenbrock, P.  Phoonthong and  F.  Renzoni, J. Mod. Opt. {\bf 58}, 1310 (2011).
\bibitem{rempe}
R.J. Thompson,  G.  Rempe,  H.J. Kimble, Phys. Rev. Lett.  {\bf 68}, 1132 (1992).
\bibitem{model1}
E.T. Jaynes,  F.W. Cummings, Proc. IEEE {\bf 51}, 89 (2005).
\bibitem{model2}
M. Tavis,  F.W.  Cummings,  Phys. Rev. {\bf 170}, 379 (1968).
\bibitem{papa}
G.J. Papadopoulos, Phys. Rev. A {\bf 37}, 2482 (1998).

\end{thebibliography}
\end{document}